\documentstyle[11pt,epsfig,epsf,axodraw]{article} 
\newcommand{\bea}{\begin{eqnarray}}
\newcommand{\eea}{\end{eqnarray}}

\begin{document}
\begin{flushright}
{\large HIP-2002-54/TH}\\ 
{hep-ph/yymmddd}\\
\end{flushright}

\begin{center}

{\Large\bf Characteristic slepton signal in anomaly mediated SUSY
breaking models via gauge boson fusion at the LHC}\\[20mm]

Anindya Datta \footnote{E-mail: datta@pcu.helsinki.fi},
Katri Huitu \footnote{E-mail: huitu@pcu.helsinki.fi} \\
{\em Helsinki Institute of Physics,\\
P.O. Box 64, 
FIN-00014 University of Helsinki, Finland}
\end{center}

\vskip 20pt
\begin{abstract}

 We point out that slepton pairs produced via gauge boson fusion in
 anomaly mediated supersymmetry breaking (AMSB) model have very
 characteristic and almost clean signal at the Large Hadron
 Collider. In this article, we discuss how one lepton associated with
 missing energy and produced in between two high-$p_T$ and high-mass
 forward jets can explore quite heavy sleptons in this scenario.

\end{abstract}

\vskip 1 true cm

Vector boson fusion (VBF) at hadronic machines such as the Large
Hadron Collider (LHC) at CERN is a useful channel for studying the
Higgs boson. Characteristic features of this mechanism are two very
energetic quark-jets, produced in the forward direction in opposite
hemispheres of the detector and carrying a large invariant mass.  The
absence of colour exchange between the forward jets ensures
suppression of hadronic activities in the central region
\cite{bjorken}.  The VBF mechanism was originally proposed to 
produce a background-free signal for a heavy Higgs \cite{dawson}.  The
usefulness of the VBF channel in uncovering an intermediate mass Higgs
has also been subsequently demonstrated \cite{zeppenfeld}.

The importance of this channel has been realized as well in searching
signals of supersymmetry (SUSY) especially in some pathological cases
\cite{datta_konar}. Following the same spirit, in this article we will
see how sleptons up to large masses can be explored via their
production by VBF and subsequent decays in the AMSB type of a model.
As sleptons are only weakly interacting, production cross-sections are
rather low.  For a minimal gravity mediated SUSY breaking (mSUGRA)
scenario, mass reach for the charged sleptons (mainly decaying to a
charged lepton and the lightest neutralino, thus producing
opposite-sign di-lepton with missing energy) at the LHC is quite low
($\sim$ 300 GeV) \cite{tata_slep}.  Due to the large production
cross-sections the strongly interacting squarks and gluinos would be
natural to investigate for the discovery of the SUSY particles
\cite{baer_tata} at the high energy hadron colliders.  As for
sleptons, couplings of the first two generations
\footnote{Throughout this paper, we will restrict ourselves to the first
two generations of the sleptons.} to $W,Z$ and
photons are almost model independent.  They are determined by the
gauge quantum numbers of these particles.  Thus 
the slepton production cross-section mainly depends
on one unknown parameter, the slepton mass.

The phenomenology of a particular SUSY model is largely determined by
the mechanism of SUSY breaking. In the following, we will see that in
the AMSB scenario, production of sleptons and their subsequent decays
can yield an almost background free signal which is very
characteristic and unique to this model. A nice feature of our
signal is that it does not depend on the input parameters of the model
in a delicate way.

Before delving into the details of signal and background, let us
discuss briefly the essence of the AMSB spectra relevant to our
analysis. A detailed description of the model and the spectra can be
found in many articles \cite{amsb,GGW,sourov}. 
Masses of
the gauginos ($M_i$) and the scalars ($M_{s}$) can be written as
the following:
\begin{equation} 
M_i = b_i\;\frac{g_i^2}{16 \pi ^2}\;m_{3/2} ;~~~
M^2_{s} = c_s\;\frac{m_{3/2}^2}{(16 \pi^2) ^2}\;+\;m_0^2 .
\label{eqn_mas}
\end{equation}

Here $b_i$'s are coefficients occurring in the $\beta$-functions of
the appropriate gauge couplings and $c_s$'s are combinations of
$\beta$-functions and anomalous dimensions of gauge and Yukawa
couplings, see {\it e.g.} \cite{GGW,sourov}. For sleptons, $c_s$'s are
negative quantities and $m_0$ is a scalar mass parameter introduced to
prevent sleptons from becoming tachyonic.  The parameter $m_0$ is the
most model dependent part of the AMSB spectrum, since there are a
number of ways to remove the tachyonic masses from the spectrum, see
{\it e.g.} \cite{fix}, leading to different additions to the scalar
masses.  We use the simplest choice, where the same mass parameter
$m_0$ is added to all the scalar masses (mAMSB model).  The gravitino
mass $m_{3/2}$ is the only other mass parameter apart from $m_0$ in
this model.

Since the gaugino masses are proportional to the beta-functions of the
corresponding gauge couplings, both the lightest neutralino (which is
the LSP) and the lighter chargino turn out to be dominated by the
wino, with their masses separated by a few hundreds of $MeV$
\cite{GGW}. The second lightest neutralino, on the other hand, is about 
three times heavier than the LSP and it is bino-dominated.  This kind
of a spectrum implies that the dominant ($\sim$ 95 \% or more, over the
whole parameter space) decay mode for the lighter chargino is $\chi^{\pm}_1
\longrightarrow \pi^{\pm} \chi^{0}_1$ \cite{CDG} . The pions in such 
cases are too soft to be detected, making the chargino practically
invisible.  The small mass difference of $\chi_1^0$ and $\chi_1^\pm$
is essential for the signal to be discussed here.  This feature is not
present in mSUGRA or gauge mediated symmetry breaking (GMSB) models,
and this is why our signal is so unique\footnote{We note that by
taking non-universal boundary conditions one can tune a model to look
like an AMSB model at any particular point of the parameter space. We
do not aim to separate between those models and AMSB.}.

Now let us focus ourselves on the sleptons. Unlike the gauginos,
slepton masses are dependent on $m_0$ as well as on $m_{3/2}$.  These
two parameters have opposite effects on the slepton mass which is
evident from eq. (\ref{eqn_mas}).  Let us
also note that the first two generations of the sleptons have almost
equal masses due to the small contributions from the corresponding
lepton masses.  In addition, the left-handed and right-handed charged
sleptons as well as the sneutrinos have masses close to each other in
the mAMSB model \cite{GGW}. Except for the stau, the
slepton masses are not very sensitive to $\tan
\beta$.

Depending on the mass parameters $m_0$ and $m_{3/2}$, the sleptons can
be heavier or lighter than the second lightest neutralino. The mass
ordering determines the decay patterns of these scalars.  
The possible decay channels of $\tilde l _L$ and $\tilde \nu$ are
\bea
\tilde l _L\rightarrow \nu_l \chi_1^- \;\; {\rm or}\;\; 
l^-\chi_i^0,\;\;\;\;
\tilde \nu_l \rightarrow l^- \chi_1^+ \;\; {\rm or}\;\; 
\nu_l\chi_i^0,
\nonumber
\eea
where $i=1,2$.  (We denote first two generations of
sleptons and leptons by $\tilde l$ and $l$.) 
Let us first
discuss the case when the sleptons are lighter than the $\chi_2 ^0$.
The decay branching
ratio to the final state comprising $\chi_1 ^\pm$ has higher (almost
the double) branching ratio than the final state comprising the LSP.
On the other hand $\tilde l _R$ can decay to $l^- \;\tau^\pm\;\tilde
\tau^\mp$ or to $l^- \chi_1 ^0$ (this two body decay can take 
place because of a tiny bino-component in $\chi_1 ^0$). The relative
strength of these decays depends on $\tan\beta$.  When kinematically
allowed, sleptons can decay to $\chi_2 ^0$.  Since $\chi_2 ^0$ is
thrice as heavy as $\chi_1 ^0$ or $\chi_1 ^\pm$, this new decay mode
of $\tilde l _L$ and $\tilde
\nu_l$ has a very small branching ratio\footnote{This branching ratio
cannot be bigger than 1- 2 $\%$ over the parameter space. So in spite
of this new decay mode, decay branching ratios to £$\chi_1 ^0$ and
$\chi_1 ^\pm$ remain practically unchanged.}.  On the other hand,
$\tilde l _R \rightarrow l^- \chi_2 ^0$ could be dominant when
kinematically allowed. It is interesting to observe that the decay
branching ratios of $\tilde l _L$ and $\tilde \nu_l$ do not change
much over the parameter space in AMSB.  This is an outcome of the 
characteristic spectrum of the model.
  $\tilde l _L$ and $\tilde \nu_l$
when decaying to the lighter chargino have a branching ratio about 66
$\%$ while rest of the time they are decaying to the lightest
neutralino.

The signals of the AMSB scenario at a future $e^+ e^-$ linear collider
were studied in the refs. \cite{sourov,roy_chou}.  However, we find it
important to look for the signature of the AMSB scenario at the LHC
which is planned to start operation in next five years. Studies to
probe anomaly mediation scenarios at the LHC, using decay cascades for
strongly interacting squarks and gluinos, have been done in
\cite{amsb_tata,barr}.  In the present work we will concentrate only
on slepton production and decay and see how our proposed signal
carries the unique stamp of this kind of a model.

We are now in a position to discuss the signal of the sleptons we are
interested in.  Since the lighter chargino is invisible in our
considerations, $\tilde l _L$ decaying to $\chi_1 ^\pm$ produces no
charged lepton and $\tilde \nu$ produces only one lepton.  Thus from
the decays of pair produced $\tilde l _L\;\tilde{\bar l _L}$ , $\tilde
\nu_l \tilde{\bar
\nu_l} $ or $\tilde l_L  \tilde{\bar \nu_L}$ one frequently has 
the {\em one lepton and missing energy} final state (with effective
branching ratios of $43 \%$, $43 \%$ and $54 \%$ respectively), unlike
in the other SUSY breaking models.  In contrast, pair production of
$\tilde l_R$s always produces pair of leptons while decaying.  We
calculate the cross-sections for the processes $pp \rightarrow {\tilde
\nu}_l {\tilde {\bar l}}_L, {\tilde \nu}_l \tilde {\bar
\nu}_l, {\tilde l}_L \tilde {\bar l}_L$ associated with two forward
jets, and consequent decays of the sleptons to produce one lepton
final state.  Therefore, the signal we will be looking for is two
high-$p_T$ jets produced in the forward region of the detector along
with an energetic lepton ($e^\pm$ or $\mu ^\pm$) produced in the
rapidity region between two jets.  One lepton final state can also
arise from the pair production of $\chi _1 ^0 \chi_2 ^0$ and
subsequent decay of $\chi_2 ^0$.  However, this contribution is
negligible compared to the ones we have discussed before \footnote{The
cross section for $pp\rightarrow \chi_1^0\chi_2^0\rightarrow$ one
lepton final state remains below $10^{-2}$ fb for the allowed
parameter range.}.

\begin{figure}[h]
{
\unitlength=1.0 pt
\SetScale{1.0}
\SetWidth{0.7}      
\scriptsize    
\noindent
\begin{picture}(95,99)(0,0)
\Text(15.0,90.0)[r]{$q$}
\ArrowLine(16.0,90.0)(58.0,90.0)
\Text(80.0,90.0)[l]{$q$}
\ArrowLine(58.0,90.0)(79.0,90.0)
\Text(53.0,80.0)[r]{$W,\gamma,Z$}
\Photon(58.0,90.0)(58.0,70.0){2}{3}
\Text(80.0,70.0)[l]{$\tilde{\ell}$, $\tilde{\nu}$ }
\DashArrowLine(79.0,70.0)(58.0,70.0){1.0}
\Text(54.0,60.0)[r]{$$}
\DashArrowLine(58.0,70.0)(58.0,50.0){1.0}
\Text(80.0,50.0)[l]{$\tilde{\ell}$, $\tilde{\nu}$ }
\DashArrowLine(58.0,50.0)(79.0,50.0){1.0}
\Text(53.0,40.0)[r]{$W,\gamma,Z$}
\Photon(58.0,50.0)(58.0,30.0){2}{3}
\Text(15.0,30.0)[r]{$q$}
\ArrowLine(16.0,30.0)(58.0,30.0)
\Text(80.0,30.0)[l]{$q$}
\ArrowLine(58.0,30.0)(79.0,30.0)
\end{picture} \ 
{} \qquad\allowbreak
\begin{picture}(95,99)(0,0)
\Text(15.0,80.0)[r]{$q$}
\ArrowLine(16.0,80.0)(58.0,80.0)
\Text(80.0,90.0)[l]{$q$}
\ArrowLine(58.0,80.0)(79.0,90.0)
\Text(53.0,70.0)[r]{$W,\gamma,Z$}
\Photon(58.0,80.0)(58.0,60.0){2}{3}
\Text(80.0,70.0)[l]{$\tilde{\ell}$, $\tilde{\nu} $}
\DashArrowLine(58.0,60.0)(79.0,70.0){1.0}
\Text(80.0,50.0)[l]{$\tilde{\ell}$, $\tilde{\nu} $}
\DashArrowLine(79.0,50.0)(58.0,60.0){1.0}
\Text(53.0,50.0)[r]{$W,\gamma,Z$}
\Photon(58.0,60.0)(58.0,40.0){2}{3}
\Text(15.0,40.0)[r]{$q$}
\ArrowLine(16.0,40.0)(58.0,40.0)
\Text(80.0,30.0)[l]{$q$}
\ArrowLine(58.0,40.0)(79.0,30.0)
\end{picture} \
{} \qquad\allowbreak
\begin{picture}(95,99)(0,0)
\Text(15.0,80.0)[r]{$q$}
\ArrowLine(16.0,80.0)(37.0,80.0)
\Text(80.0,90.0)[l]{$q$}
\ArrowLine(37.0,80.0)(79.0,90.0)
\Text(33.0,70.0)[r]{$W$}
\Photon(37.0,80.0)(37.0,60.0){2}{3}
\Text(49.0,65.0)[b]{$\gamma,Z$}
\Photon(37.0,60.0)(58.0,60.0){2}{3}
\Text(80.0,70.0)[l]{$\tilde{\ell}$, $\tilde{\nu}$}
\DashArrowLine(58.0,60.0)(79.0,70.0){1.0}
\Text(80.0,50.0)[l]{$\tilde{\ell}$,  $\tilde{\nu}$}
\DashArrowLine(79.0,50.0)(58.0,60.0){1.0}
\Text(33.0,50.0)[r]{$W$}
\Photon(37.0,60.0)(37.0,40.0){2}{3}
\Text(15.0,40.0)[r]{$q$}
\ArrowLine(16.0,40.0)(37.0,40.0)
\Text(80.0,30.0)[l]{$q$}
\ArrowLine(37.0,40.0)(79.0,30.0)
\end{picture} \ 
{} \qquad\allowbreak
\begin{picture}(95,99)(0,0)
\Text(15.0,80.0)[r]{$q$}
\ArrowLine(16.0,80.0)(37.0,80.0)
\Text(80.0,90.0)[l]{$q$}
\ArrowLine(37.0,80.0)(79.0,90.0)
\Text(33.0,70.0)[r]{$W,Z$}
\Photon(37.0,80.0)(37.0,60.0){2}{3}
\Text(50.0,63.0)[b]{$h,H$}
\DashLine(37.0,60.0)(58.0,60.0){3.0}
\Text(80.0,70.0)[l]{$\tilde{\ell}$, $\tilde{\nu}  $}
\DashArrowLine(58.0,60.0)(79.0,70.0){1.0}
\Text(80.0,50.0)[l]{$\tilde{\ell}$,  $\tilde{\nu}$ }
\DashArrowLine(79.0,50.0)(58.0,60.0){1.0}
\Text(33.0,50.0)[r]{$W,Z$}
\Photon(37.0,60.0)(37.0,40.0){2}{3}
\Text(15.0,40.0)[r]{$q$}
\ArrowLine(16.0,40.0)(37.0,40.0)
\ArrowLine(37.0,40.0)(79.0,30.0)
\Text(80.0,30.0)[l]{$q$}
\end{picture} \ 
}
{
\unitlength=1.0 pt
\SetScale{1.0}
\SetWidth{0.7}      
\scriptsize    
\noindent
\begin{picture}(95,99)(0,0)
\Text(15.0,90.0)[r]{$g$}
\Text(15.0,50.0)[r]{$g$}
\Gluon(16.0,90.0)(58.0,90.0){4}{3}
\ArrowLine(58.0,90.0)(89.0,98.0)
\Text(60.0,78.0)[l]{$W,\gamma$}
\Text(60.0,62.0)[l]{$Z$}
\Gluon(16.0,50.0)(58.0,50.0){4}{3}
\Text(80.0,33.0)[l]{$q$}
\Text(80.0,103.0)[l]{$q$}
\ArrowLine(58.0,50.0)(89.0,42.0)
\ArrowLine(58.0,90.0)(58.0,50.0)
\Photon(58.0,70.0)(78.0,70.0){2}{3}
\DashArrowLine(78.0,70.0)(100.0,82.0){1.0}
\DashArrowLine(78.0,70.0)(100.0,58.0){1.0}
\Text(100.0,80.0)[l]{$\tilde{\ell}$, $\tilde{\nu}$ }
\Text(100.0,60.0)[l]{$\tilde{\ell}$, $\tilde{\nu}$ }
\end{picture} \ 
{} \qquad\allowbreak
\begin{picture}(95,99)(0,0)
\Text(15.0,90.0)[r]{$q$}
\Text(15.0,50.0)[r]{$q$}
\ArrowLine(16.0,90.0)(58.0,90.0)
\ArrowLine(58.0,90.0)(89.0,98.0)
\ArrowLine(16.0,50.0)(58.0,50.0)
\Text(90.0,40.0)[l]{$q$}
\Text(90.0,100.0)[l]{$q$}
\ArrowLine(58.0,50.0)(89.0,42.0)
\Gluon(58.0,90.0)(58.0,50.0){3}{4}
\Photon(72.0,45.0)(78.0,70.0){2}{3}
\DashArrowLine(78.0,70.0)(100.0,82.0){1.0}
\DashArrowLine(78.0,70.0)(100.0,58.0){1.0}
\Text(105.0,80.0)[l]{$\tilde{\ell}$, $\tilde{\nu}$ }
\Text(105.0,60.0)[l]{$\tilde{\ell}$, $\tilde{\nu}$ }
\Text(50.0,70.0)[l]{$g$}
\Text(80.0,51.0)[l]{$W,\gamma, Z$}
\end{picture} \ 
{} \qquad\allowbreak
\begin{picture}(95,99)(0,0)
\Text(15.0,90.0)[r]{$q$}
\Text(15.0,50.0)[r]{$q$}
\Text(60.0,78.0)[l]{$W,Z$}
\Text(60.0,61.0)[l]{$g ,\gamma$}
\Text(105.0,97.0)[r]{$q$}
\Text(105.0,45.0)[r]{$q$}
\Text(120.0,76.0)[l]{$\tilde{\ell}$, $\tilde{\nu}$ }
\Text(120.0,34.0)[l]{$\tilde{\ell}$, $\tilde{\nu}$ }
\ArrowLine(16.0,90.0)(50.0,70.0)
\ArrowLine(16.0,50.0)(50.0,70.0)
\Photon(50.0,70.0)(80.0,70.0){3}{4}
\ArrowLine(80.0,70.0)(104.0,94.0)
\ArrowLine(80.0,70.0)(104.0,44.0)
\Photon(93.0,55.0)(110.0,55.0){2}{3}
\DashArrowLine(110.0,55.0)(125.0,70.0){1.0}
\DashArrowLine(110.0,55.0)(125.0,40.0){1.0}
\end{picture} \
{} \qquad\allowbreak
\begin{picture}(95,99)(0,0)
\Text(15.0,90.0)[r]{$q$}
\Text(15.0,50.0)[r]{$q$}
\Text(60.0,78.0)[l]{$W,Z$}
\Text(60.0,61.0)[l]{$\gamma$}
\Text(105.0,97.0)[r]{$\tilde{\ell}$, $\tilde{\nu}$}
\Text(105.0,45.0)[r]{$\tilde{\ell}$, $\tilde{\nu}$}
\Text(60.0,97.0)[l]{$g$}
\Text(60.0,35.0)[l]{$g$}
\ArrowLine(16.0,90.0)(50.0,70.0)
\ArrowLine(16.0,50.0)(50.0,70.0)
\Photon(50.0,70.0)(80.0,70.0){3}{4}
\DashArrowLine(80.0,70.0)(104.0,94.0){1.0}
\DashArrowLine(80.0,70.0)(104.0,44.0){1.0}
\Gluon(34.0,80.0)(58.0,100.0){3}{4}
\Gluon(34.0,60.0)(58.0,40.0){3}{4}
\end{picture} \
}

\caption{\em Generic parton level diagrams leading to slepton
  pair production with two jets at hadronic colliders. The diagrams
in the first row dominantly contribute to our considered signal. 
Diagrams in the second row have a small contribution to our signal
due to kinematic cuts. 
}
\label{diagrs}
\end{figure}
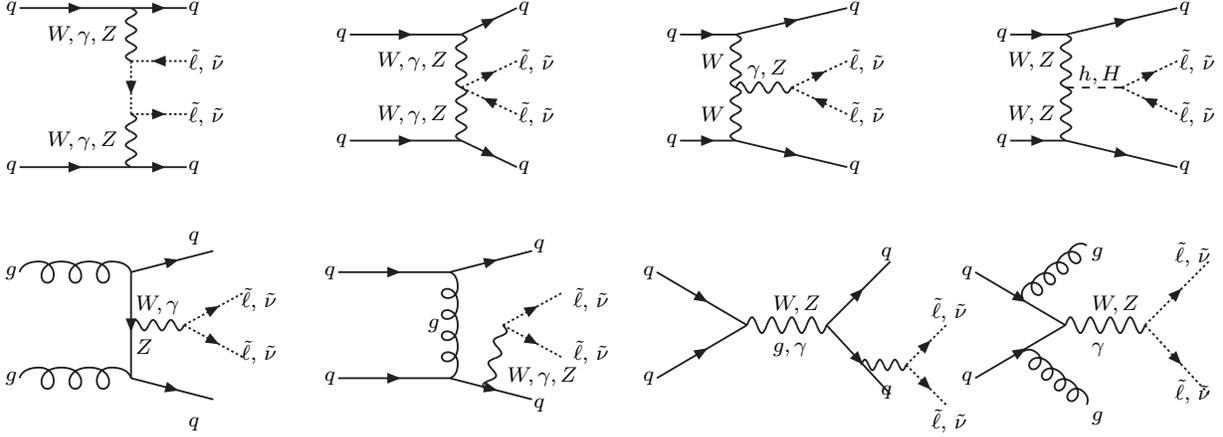

The relevant generic Feynman diagrams are depicted in
fig. \ref{diagrs}.  In the first row we have presented those diagrams
which are really due to vector boson fusion. These diagrams contribute
to the cross-section dominantly.  The diagrams in the second row
contribute to the slepton pair production along with two jets, but we
have explicitly checked that they do not contribute to the total
cross-section significantly as they do not survive the kinematic cuts.
It is also important to note that for the first two diagrams in the second
row, colored particles are exchanged along two final partons. For
these processes, there are gluon emissions in the central rapidity
region.  As we demand for our signal that the central region between
two forward jets be hadronically quiet, we expect that these
diagrams would not pass the central jet veto. However, since we are
limited to the LO diagrams, simulation of jet emission in the central
region and use of central jet veto is beyond our calculation.  We have
evaluated numerically the amplitudes using the computer code HELAS
\cite{HELAS} and then used a Monte Carlo routine to integrate over the
phase-space.  In our numerical calculations we have used the CTEQ4L
set of parton parametrisation in \cite{cteq}.

\begin{figure}[t]
\vspace*{-5em}
\centerline{\hspace*{3em}
\epsfxsize=9cm\epsfysize=8.0cm
                     \epsfbox{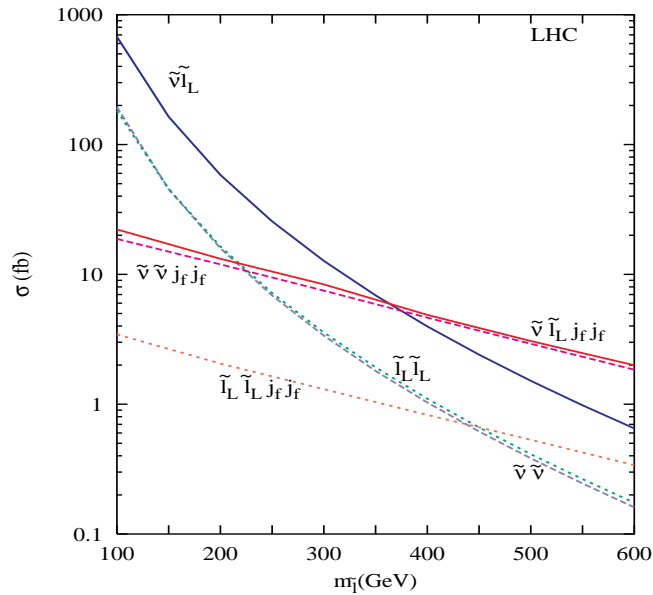}
}
\caption[]{\em Cross-sections of slepton pair production from VBF and DY channels. 
Kinematic cuts defined in the text have been used to calculate the 
VBF cross-sections. }
\label{cross}
\end{figure}

To show the efficacy of our chosen VBF channel for large slepton
masses, we will first present the cross-sections for the slepton pair
production at the LHC energies in fig. \ref{cross}. For the purpose of
comparison, we will also present the cross-section of slepton pair via
Drell-Yan (DY) production. From the figure it is evident that for
large slepton masses the VBF cross-section is larger than the DY
cross-section, which falls off quite fast with slepton mass.  The
advantage of using the VBF method for slepton production at LHC
becomes even more obvious - at least at the preliminary stage - when
we note that in calculating the VBF cross-sections we have used the
kinematic cuts defined in the following part of the text, while there
are no kinematic cuts applied in calculation of the DY cross-section.

The proposed signal is not background free.  Similar events can be
faked by the production of a real or virtual $W$ along with two
forward jets, followed by the leptonic decay of the $W$. Such final
states can arise in the SM from real emission corrections to single
$W$-production process (${\cal O} (\alpha _s ^2 \alpha_W ^2)$) as well
as from the electroweak $W$ production along with two jets (${\cal
O}(\alpha_W ^4)$). We have calculated the full tree-level contribution
to the process $pp \rightarrow j_f j_f (W,W^\ast \rightarrow) l \nu$
using the package MADGRAPH \cite{madgraph}.  Transverse mass
distribution of the above background has a sharp edge for the
leptons coming from the decay of a real W. On the other hand, $l \nu$
coming from a $W^\ast$, can smear the sharp edge of this
transverse-mass distribution. The finite detector resolution is
another source for smearing.
To take into account the detector effects we have (gaussian) smeared 
the jet and lepton
energies using \cite{smear}:
\begin{center}
$\Delta E_j / E_j =  0.6/\sqrt{E_j} + 0.03, 
~~~~~\Delta E_l / E_l =  0.15/\sqrt{E_j} + 0.01$ 
\end{center}
\noindent
for the signal and all background processes. Finally, we calculate the 
missing $p_T$ by imposing the conservation of momentum along transverse
direction.  

\begin{figure}[t]
\vspace*{-5em}
\centerline{\hspace*{3em}
\epsfxsize=9cm\epsfysize=8.0cm
                     \epsfbox{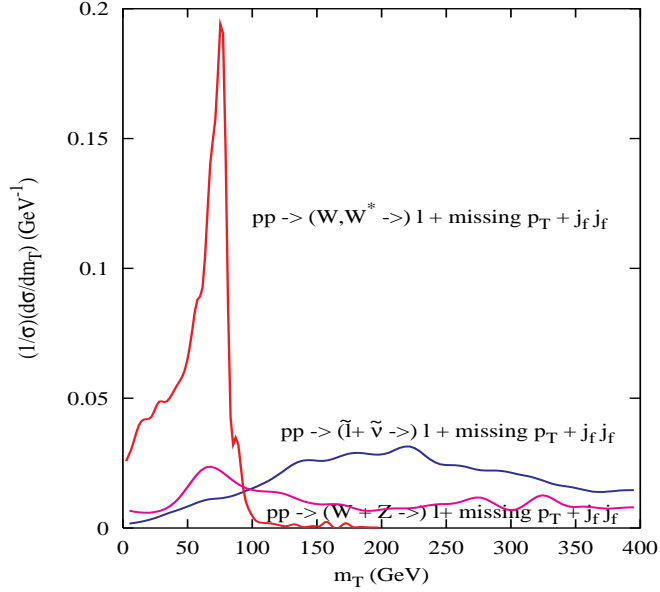}
}
\caption[]{\em Normalised transverse mass distribution of signal and backgrounds. 
The small structures in the plots are due to  Monte Carlo fluctuations.}
\label{transmass}
\end{figure}

The other source of one lepton background is $WZ$ production along
with 2 forward jets, followed by the leptonic decay of $W$ and
invisible decay of $Z$ to a pair of neutrinos. We have also estimated this
background. 

We plot the normalised transverse mass distribution of signal (coming
from $\tilde l \tilde \nu$ production and decay. For this plot we have
used $m_{\tilde l} = m_{\tilde \nu}= 163$ GeV and $m_{\chi^0_1} =
m_{\chi^+_1} = 90$ GeV. The kinematics of all three processes
contributing to the signal are very similar, since all the sleptons as
well as the lightest neutralino and chargino are almost degenerate in
mass in AMSB model. Thus we present only the $\tilde l \tilde \nu$
case.) and backgrounds in figure \ref{transmass} for illustration.  The
sharp decrease of $W/W^\ast$ background around W-mass, clearly points
to the fact that dominant part of this background comes from a real
$W$ production, while the tail of this distribution beyond W-mass is
due to the contribution from off-shell $W$, in addition to the detector
resolution effects, which we have taken into account as well. 
The other background
from $WZ$ production has similar kind of transverse mass distribution
to that of the signal. For heavier charged-sleptons/sneutrinos the
maximum (which is not very sharp) of the signal distribution is shifted
more towards right. From the distributions it is evident that once we
demand that the transverse mass constructed from the lepton and the
missing $p_T$ vector is greater than 100 GeV, a large part of
$W,W^\ast$ background can be removed.  In this process we loose part of the
signal and $WZ$ background. We will soon see that the remaining number
of signal events is enough to be statistically significant over the
remaining background.

We have used the following set of cuts to minimize the background:\\[4pt]
\hspace*{0.5cm}$\bullet$ Two forward jets in opposite hemispheres, 
      with $E_T > 40~GeV$ and 
      $2.0 \le |\eta_j| \le 5.0$. \\
\hspace*{0.5cm}$\bullet$ $\Delta \eta_{jj} > 4$.\\
\hspace*{0.5cm}$\bullet$ $M_{inv} (jj) > 650~GeV$.\\
\hspace*{0.5cm}$\bullet$ ${E_T}\!\!\!\!\!\!\!\slash \;\;> 100~GeV$.\\
\hspace*{0.5cm}$\bullet$ ${p_T ^l} > 15~GeV$.\\
\hspace*{0.5cm}$\bullet$ $|\eta_l| < 2$.\\
\hspace*{0.5cm}$\bullet$ ${m_T}(l{p_T}\!\!\!\!\!\!\slash\;\;)  >
100~GeV$.
\vspace*{4pt}

Use of the above set of cuts reduces the one lepton background down to
16 fb (14 fb from $WZ$ and 2 fb from $W/W^\ast$ production).  We have
also estimated backgrounds coming from $W^+W^- + $ 2 forward jets
$\rightarrow \nu\nu l^+l^- +$ 2 forward jets and $(\gamma ^{\ast}, Z
\rightarrow) l^+l^- +$ 2 forward jets production, with one lepton
going down the beam pipe.  After the cuts these backgrounds are less
than $10^{-5}$ fb.

To calculate the signal significance over the background, it is very
important to know the normalisation of the latter
accurately. Unfortunately, the normalisation for the $WZ$ production
associated with two forward jets is not known. One of the reasons is
that we require that there is no jet activity in the central part of
the detector, and to parametrise this kind of veto is beyond the
perturbative QCD calculations.  These issues have been discussed in
detail in ref. \cite{eboli}. Thus the exact normalisation has to be
calculated from the LHC data itself.  We just want to comment that
this is possible in principle by estimating $pp \rightarrow Z
(\rightarrow e^+ e^-) W (\rightarrow \mu \nu_\mu) +$ {\it 2 forward
jets} events.  From this three lepton final state we can get
cross-section for one lepton final state by scaling with the
appropriate branching ratios.  Since the background cross-section is
only of the leading order it depends strongly on the choice of the
scale of $\alpha _s$ and also on the factorization scale of the parton
distribution functions. We have chosen the scale to be at min ($p_T
^{j_1}, p_T ^{j_2}$) when estimating the background. On the other hand
for the signal the factorization scale for parton distribution
functions has been chosen to be at the sum of the slepton masses we
are producing.

\begin{figure}[t]
\vspace*{-5em}
\centerline{\hspace*{3em}
\epsfxsize=9cm\epsfysize=8.0cm
                     \epsfbox{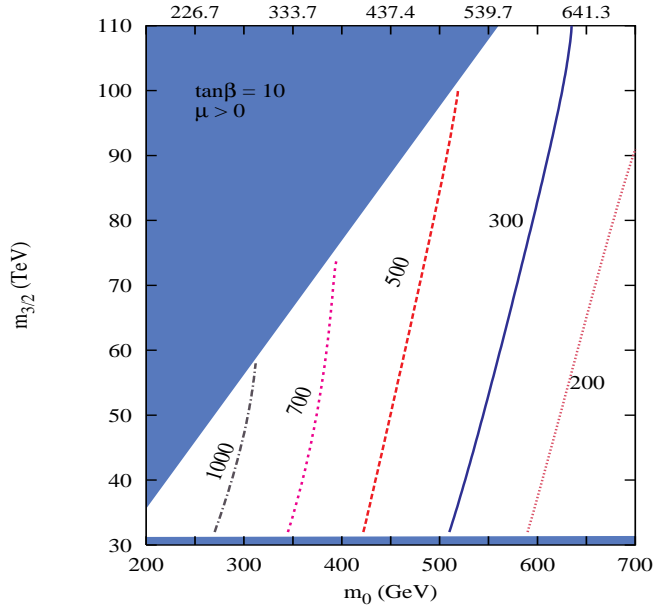}
}
\caption[]{\em Lines of constant number of events in $m_0 - m_{3/2}$ plane for 
$\tan \beta = 10$ and $\mu > 0$. The shaded regions are constrained 
from  theoretical and experimental considerations. On the upper x-axis
the $\tilde l_L$ masses are given for $m_{3/2}$ = 30 TeV.
We have assumed 100$fb^{-1}$ for integrated luminosity.}
\label{evts}
\end{figure}

We present the numbers of 1-lepton + ${E_T}\!\!\!\!\!\!\!\slash
\;$ + 2 forward-jets events in Fig. \ref{evts}. We choose to present 
the contours of constant number of events in $m_0 - m_{3/2}$ plane for
$\tan \beta = 10$ and $\mu > 0$. Results are not very sensitive to the
sign of $\mu$. As mentioned earlier, the
interactions necessary to calculate the slepton pair production in VBF
do not depend on any unknown SUSY parameters.  As we are interested
only in the first two generations of the sleptons, mixing of the left-
and right-sleptons is not very important. One also needs the couplings
of CP-even neutral Higgses to a pair of sleptons (which are somewhat
model dependent) but we checked that those diagrams contribute little
to the total cross-section. Thus, slepton pair production
cross-section along with two forward jets depends mainly on the
slepton masses. Increasing $m_0$ will decrease the cross-section,
while increasing $m_{3/2}$ tends to increase the cross-section by
decreasing the slepton masses. Let us now come to the decay of the
sleptons. It has already been pointed out that changing the input
parameters has little impact on the slepton branching ratios unless we
are close to the kinematic limits.  On the other hand, the decay
kinematics depends somewhat on the parameters.  As we increase
$m_{3/2}$, the masses of $\chi_1 ^0$ and $\chi_1^\pm$ increase and
affect the two body decay kinematics. The resulting contours, for the
number of events, closely resemble the contours of constant slepton
masses. The $\tilde e_L$ masses for $m_{3/2} = 30$ TeV are also shown
in the plot on the upper x-axis to indicate how the production
cross-section falls with slepton mass. We want to point out that, from
the Figure \ref{evts}, it is evident that for large slepton masses the
cross-section falls off quite slowly.  This is in contrast with the
Drell-Yan production of sleptons
\cite{baer_reno}. One can easily check from the plot that production 
cross-section (${\tilde \nu}_l {\tilde {\bar l}}_L + {\tilde \nu}_l
\tilde {\bar \nu}_l + {\tilde l}_L \tilde {\bar l}_L$) is at the fb
level for 500 GeV slepton masses after the suppression from branching
ratios and kinematic cuts, while in the Drell-Yan case
\cite{baer_reno} the raw cross-section is of the order of 1 fb for a
slepton mass of 500 GeV. This clarifies why we stick to the VBF
channel for producing the slepton pairs. Indeed, for low mass
sleptons, cross-section from direct production \cite{baer_reno} is
always an order of magnitude higher. However, the direct production
cross-section falls quite fast with slepton mass. For slepton masses
greater than 450 GeV, the VBF cross-section is larger than the direct
one.

With 100 $fb ^{-1}$ integrated luminosity, 16 $fb$ of the SM
background implies that one needs to have around 200 signal events for
a 5$\sigma$ discovery and around 120 events for a 3$\sigma$ discovery.
It is evident from the figure \ref{evts}, that one can probe slepton masses up
to 600 GeV which is quite remarkable. If we assume a $5 \%$
uncertainty in our estimated background (as we do not know the
background normalisation accurately) and add this error with our
estimated number of background events in quadrature \cite{eboli}, the above
5$\sigma$ mass limit goes down close to 450 GeV (in this case one needs
nearly 450 events for the 5$\sigma$ discovery).
 
One could have one lepton signal in mSUGRA (and GMSB 
\footnote{We note that in the simplest GMSB model, over most of
the parameter space $\chi_1^0$ decays dominantly to a photon and a
gravitino.  This provides an additional method to discriminate between
GMSB and AMSB.
}) model as well. The dominant
contribution comes from the production of $\tilde l_L (\rightarrow l
\chi_1^0) \tilde \nu (\rightarrow 
\nu \chi_1^0)$ and $\chi_1^\pm (\rightarrow l \nu\chi_1^0)\;\chi_2^0
(\rightarrow \nu \nu \chi_1^0)$ associated with two forward jets.  The
cross-section can be as large as 6 fb for the former one for light
sleptons, while the latter one can be of the order of 0.1 fb.
The sum is comparable with the one lepton cross section in the AMSB
scenario. Thus one might assume that our proposed signal cannot be used as
some benchmark for this scenario. However, neutralinos
and charginos in mSUGRA or GMSB, have a very distinct and clean $3l +
E_T \!\!\!\!\!\!\!/\;\;$ signature coming from $pp \rightarrow \chi_1^\pm
(\rightarrow l \nu\chi_1^0)\;\chi_2^0 (\rightarrow l \bar l
\chi_1^0)$ \cite{tri-lep}. 
(This cross-section can be as high as 100 fb in regions 
of parameter space.) This is practically absent in the AMSB scenario.
We have explicitly checked that in AMSB, the cross section 
of $3l + E_T \!\!\!\!\!\!\!/\;\;$ final state from chargino
neutralino and slepton production and decay
is always below $10^{-2}$ fb level. The absence of this clean
tri-lepton final state but appearance of 1 lepton + 2 forward jet
signature clearly points to the 'AMSB' kind of SUSY breaking.

Moreover it is important to note that in the AMSB-like scenarios, lighter
chargino has a comparatively long life time (unlike in mSUGRA and
GMSB), and it may leave an
ionized track in the detector before its decay. This issue has been
discussed and exploited to dig out the signals of AMSB  
\cite{barr, feng}. In our analysis we have assumed that the
chargino is almost invisible, but detailed simulation in the context of
AMSB search at LHC reveals \cite{barr} that identifying this kind of a
macroscopic ionizing tracks is possible with high efficiency
with off-line analysis of the data. In that case, our signal will
indeed be free of any 'physics-background' and the mass reach will
certainly improve beyond 600 GeV.

The shaded region parallel to the $x$-axis in the plot is ruled out from the
direct chargino search at LEP \cite{lep_data}. Region shaded in the
upper left corner of the figure is excluded from the consideration
that in this region the lighter stau is the lightest supersymmetric
particle which may not be desirable from cosmological
considerations. There are also other constraints coming from the
considerations of radiative decay of B-mesons and muon $(g -2)$
\cite{bsg_mu_ufb_amsb}, and also from analyzing the nature of the
vacuum due to the scalar sector of AMSB model
\cite{ufb_amsb}. We have not shown or considered them, but the 
effect of the constraints on the 
parameter space can be readily read from the given references.
However, direct
search experiments should not be biased by other experimental
results, which may, furthermore, be model dependent.
We want to stress that our proposed signal is unique and can
provide a smoking gun signature of this kind of SUSY breaking at a
hadronic collider like LHC over a large area of parameter space.

Furthermore, we emphasize that the signal we propose is
characteristic of not only the minimal AMSB scenario but also to the
other variants of this model, or any model with AMSB-like mass
spectrum.  The one lepton signature is an outcome of the invisibility
of the chargino decay and mass spectrum of the model.  Decay pattern
of the $\chi_1 ^\pm$, along with its composition (also the composition
of $\chi_1 ^0$) are general features of the considered alternative
AMSB models. 
In several variants of the AMSB type models,
for the
first two generations of sleptons ${\tilde {l}}_L, {\tilde \nu}_l$
masses are not only close to each other, but they are heavier than the
$\chi_1 ^\pm$ or $\chi_1 ^0$.
This ensures
the previously discussed decay pattern and decay branching ratios for
these particles.

To summarise, pair production of sleptons associated with two forward
jets, and their decay to one lepton final state can be used at the LHC
to look for a signal of anomaly mediated SUSY breaking model. We have
analyzed the signal and background and have shown that up to large
slepton masses one can have a very distinct signature of this model
coming from the slepton pair production and decay.  Our proposed
signal not only characterizes the minimal model but also the other
variants of AMSB.

{\bf Acknowledgments:} Authors thank the Academy
of Finland (project number 48787) for financial support.
 
\newcommand{\plb}[3]{{Phys. Lett.} {\bf B#1} #2 (#3)}                  %
\newcommand{\prl}[3]{Phys. Rev. Lett. {\bf #1} #2 (#3) }        %
\newcommand{\rmp}[3]{Rev. Mod.  Phys. {\bf #1} #2 (#3)}             %
\newcommand{\prep}[3]{Phys. Rep. {\bf #1} #2 (#3)}                   %
\newcommand{\rpp}[3]{Rep. Prog. Phys. {\bf #1} #2 (#3)}             %
\newcommand{\prd}[3]{Phys. Rev. {\bf D#1} #2 (#3)}                    %
\newcommand{\np}[3]{Nucl. Phys. {\bf B#1} #2 (#3)}                     %
\newcommand{\npbps}[3]{Nucl. Phys. B (Proc. Suppl.)
           {\bf #1} #2 (#3)}                                           %
\newcommand{\sci}[3]{Science {\bf #1} #2 (#3)}                 %
\newcommand{\zp}[3]{Z.~Phys. C{\bf#1} #2 (#3)}  
\newcommand{\epj}[3]{Eur. Phys. J. {\bf C#1} #2 (#3)} 
\newcommand{\mpla}[3]{Mod. Phys. Lett. {\bf A#1} #2 (#3)}             %
 \newcommand{\apj}[3]{ Astrophys. J.\/ {\bf #1} #2 (#3)}       %
\newcommand{\jhep}[2]{{Jour. High Energy Phys.\/} {\bf #1} (#2) }%
\newcommand{\astropp}[3]{Astropart. Phys. {\bf #1} #2 (#3)}            %
\newcommand{\ib}[3]{{ ibid.\/} {\bf #1} #2 (#3)}                    %
\newcommand{\nat}[3]{Nature (London) {\bf #1} #2 (#3)}         %
 \newcommand{\app}[3]{{ Acta Phys. Polon.   B\/}{\bf #1} #2 (#3)}%
\newcommand{\nuovocim}[3]{Nuovo Cim. {\bf C#1} #2 (#3)}         %
\newcommand{\yadfiz}[4]{Yad. Fiz. {\bf #1} #2 (#3);             %
Sov. J. Nucl.  Phys. {\bf #1} #3 (#4)]}               %
\newcommand{\jetp}[6]{{Zh. Eksp. Teor. Fiz.\/} {\bf #1} (#3) #2;
           {JETP } {\bf #4} (#6) #5}%
\newcommand{\philt}[3]{Phil. Trans. Roy. Soc. London A {\bf #1} #2
        (#3)}                                                          %
\newcommand{\hepph}[1]{hep--ph/#1}           %
\newcommand{\hepex}[1]{hep--ex/#1}           %
\newcommand{\astro}[1]{(astro--ph/#1)}         %

\end{document}